# Denoising Magnetic Resonance Spectroscopy (MRS) Data Using Stacked Autoencoder for Improving Signal-to-Noise Ratio and Speed of MRS

Jing Wang[1], Bing Ji[2], Yang Lei[1], Tian Liu[3], Hui Mao[2*], Xiaofeng Yang[1*]

[1]Department of Radiation Oncology and Winship Cancer Institute, Emory University, Atlanta, GA

[2]Department of Radiology and Imaging Science and Winship Cancer Institute, Emory University, Atlanta, GA

[3]Department of Radiation Oncology, Icahn School of Medicine at Mount Sinai, New York, NY

*Corresponding to: xiaofeng.yang@emory.edu and hmao@emory.edu

**Running title**: MRS data denoising

**Manuscript Type**: Original Research

# ABSTRACT

**Background**: While magnetic resonance imaging (MRI) provides high resolution anatomical images with sharp soft tissue contrast, magnetic resonance spectroscopy (MRS) enables non-invasive detection and measurement of biochemicals and metabolites. However, MRS has low signal-to-noise ratio (SNR) when concentrations of metabolites are in the range of the million molars. Standard approach of using a high number of signal averaging (NSA) to achieve sufficient NSR comes at the cost of a long acquisition time.

**Purpose:** We propose to use deep-learning approaches to denoise MRS data without increasing the NSA. This method has the potential to reduce acquisition times as well as improve SNR and quality of MRS data which could ultimately enhance the diagnostic value and broaden the clinical applications of MRS.

**Methods:** The study was conducted using data collected from the brain spectroscopy phantom and human subjects. We utilized a stack auto-encoder (SAE) network to train deep learning models for denoising low NSA data (NSA = 1, 2, 4, 8, and 16) randomly truncated from high SNR data collected with high NSA (NSA=192) which were also used to obtain the ground truth. We applied both self-supervised and fully-supervised training approaches and compared their performance of denoising low NSA data based on improved SNRs. To prevent overfitting, the SAE network was trained in a patch-based manner. We then tested the denoising methods on noise-containing data collected from the phantom and two human subjects. We evaluated their performance by comparing the SNR levels and mean squared errors (MSEs) calculated for the whole spectra against high SNR "ground truth", as well as the value of chemical shift of N-acetyl-aspartate (NAA) before and after denoising.

**Results**: With the SAE model, the SNR of low NSA data (NSA = 1) obtained from the phantom increased by 22.8% and the MSE decreased by 47.3%. For low NSA images of the human parietal and temporal lobes, the SNR increased by 43.8% and the MSE decreased by 68.8%. In all cases, the chemical shift of NAA in the denoised spectra closely matched with the high SNR spectra, suggesting no distortion to the spectra from denoising. Furthermore, the denoising performance of the SAE model was more effective in denoising spectra with higher noise levels.

**Conclusions**: The reported SAE denoising method is a model-free approach to enhance the SNR of low NSA MRS data. With the denoising capability, it is possible to acquire MRS data with a few NSA, resulting in shorter scan times while maintaining adequate spectroscopic information for detecting and quantifying the metabolites of interest. This approach has the potential to improve the efficiency and effectiveness of clinical MRS data acquisition by reducing scan time and increasing the quality of spectroscopic data.

# 1. INTRODUCTION

Magnetic resonance spectroscopy (MRS) is a clinically available non-invasive analytical tool to obtain biochemical and metabolic information of tissue *in vivo*[1-4]. When combined with magnetic resonance imaging (MRI), it can be used to directly detect and measure concentrations of metabolites in the sampling volume well-defined by MRI[5,6]. MRS has been shown to be applicable and beneficial in clinical radiological exams, particularly in the areas of characterizations of brain tumors[7,8], prostate[9,10] and breast cancers[11,12] as well as some neurological diseases, such as epilepsy[13,14]. However, MRS signals from the metabolites of interest are usually orders of magnitude lower than the water signal used in conventional MRI[15,16]. This intrinsic limitation of low signal-to-noise ratio (SNR) hinders the clinical applications of MRS[17]. Although SNR can be increased by acquiring data using a large sample volume or a high number of signal averages (NSA), these approaches have inevitable limitations in *in vivo* applications, such as poor spatial resolution and low specificity when sampling the heterogeneous tissue and long data acquisition time. Furthermore, analyzing and interpreting spectroscopic data heavily depend on experience and prior knowledge, making data processing difficult to adopt in the radiology workflow. Therefore, integrating MRS into clinical applications to gain metabolic information and biomarkers for improving individualized diagnosis has been largely overlooked[18].

Given the increasing need in molecular and metabolic imaging and quantitative assessment of the diseases in precision medicine, there has been renewed interest in broadening and improving clinical applications of MRS. Efforts to accelerate MRS data acquisition include the development of ultrafast acquisition methods to obtain high-resolution MRS imaging[19], utilization of ultra-high field systems or parallel data acquisition, and implementation of data-driven reconstructions[20]. However, one of the strategies to improve the sensitivity of MRS with existing instrumentation capabilities at the current clinical field strength (*i.e.,* 3 Tesla) is to reduce or even completely remove noises while retaining the signal as a part of the post-processing step. Traditionally, noises can be mitigated by applying a specific form of filters, electronically or digitally, with the penalty of sacrificing portions of signals due to incomplete separation of signals and noise, especially when noises overlapping on signals. More advanced denoising approaches, such as the wavelet thresholding[21,22] and wavelet shrinkage[23,24] methods or most recently, machine-learning assisted wavelet feature analysis and classification methods[25,26], were developed to distinguish the signal from the noise before removing the noise components to improve the SNR. However, these denoising approaches are mostly based on prior knowledge to describe the forms and features of the noise or signal, which may not capture all the noise patterns encountered in clinical settings.

Here, we present a data-driven deep learning-based spectral denoise method for enhancing SNR of MRS data. The stacked autoencoder (SAE) network with local denoising and feature learning capabilities[27,28] has been used to denoise "spectrum-like" electrocardiogram (ECG) data[29-32]. SAE uses an encoder to learn the higher-level hierachical features and a decoder to reconstruct those features back to original-sized signals, while discarding the noisy components embedded in the original input. Like those of ECG data, noise peaks in MRS data MRS have sparse representations in the feature vector domain. Thus, we can identify and characterize them using a patch-based SAE network. This approach has two distinctive strengths: 1) it uses a patch-based method to enlarge data variation when denoising the spectroscopic data, and 2) it is not dependent on prior knowledge for the sparse representation. For applications in denoising MRS data, SAEs were trained with both fully-supervised and self-supervised manners. Our results demonstrated that this new approach can effectively reduce the noise level of low NSA spectra collected in a short acquisition time (NSA = 4-8 and < 16 seconds), obtaining comparable metabolite signals to those from long scans, *e.g.*, ~6-7 minutes.

## 2. MATERIALS AND METHODS

### 2.1. Acquisition of MRS data

High SNR data were collected from a brain MRS phantom (Model 2152220, GE Healthcare, Chicago, IL) and healthy human subjects (N=2) using a 3T MR scanner (MAGNETOM Prisma, Siemens Healthcare, Erlangen, Germany). The phantom contains metabolites with known concentrations, including N-acetylaspartate (NAA, 12.5 mM), creatine (Cre, 10 mM), choline (Cho, 3 mM), lactate (Lac, 15 mM), glutamate (Glu, 12.5 mM) and myo-inositol (MI, 7.5 mM). T1-weighted magnetization-prepared rapid acquisition gradient echo (MPRAGE) images were recorded first with TR of 500 ms; TE of 8.7 ms; flip angle of 90°, matrix of 180×240 mm$^2$, field of view (FOV) of 192×256 pixels, slice thickness of 2 mm to generate T1 weighted images in three different orthogonal directions. A sampling voxel of 20×20×20 mm$^3$ then was placed at the position based on the images. For the data collected from the phantom, we selected eight different locations to count for the location-dependent variations in the training dataset, including the anterior down (A_DN), anterior high (AH), front left (FL), front right (FR), high (H), posterior left (PL), posterior right (PR), and posterior superior (PS) in reference to the isocenter, with examples shown in **Figure 1 (a, b)**. For human subjects, we collected data from a single voxel placed in the parietal and temporal brain regions, as depicted in **Figure 1(c, d)**. Single-voxel proton ($^1$H) spectroscopy (SVS) data were collected using a point resolved spectroscopy sequence (PRESS) with an echo time (TE) of 30 ms, a repetition time (TR) of 2000 ms, an acquisition bandwidth of 1200 Hz and a vector size of 1024, with a standard shimming and water suppression methods implemented on the scanner. The total MRS acquisition times, excluding pre-acquisition shimming and water suppression, were 6 minutes and 40 seconds for the high NSA spectra, including 8 pre-scans. Phase cycling of 4 was used for both phantom and human data acquisition when NSA is over 4. The data were saved individually at each NSA, so that 192 individual spectra as well as an overall spectrum averaged from 192 individual ones were separately stored and can be retrieved for analysis. More details of acquisition protocols of MRS data can be found in our previous work[33]. The human subject study was conducted in accordance with the institutional review board, and written consents were obtained from all participants.

To test the developed method in a clinically relevant application, we retrospectively performed spectral denoising using low SNR data extracted from spectroscopic images acquired from two brain tumor patients. Briefly, individual spectra were extracted from the multivoxel data of chemical shift imaging (CSI) with a FOV of 16×16 voxels collected with NSA = 1. The details of obtaining the low SNR dataset from CSI data have been reported in the previous study[33]. Spectra extracted from the CSI have intrinsically low SNR due to low NSA. As these are low SNR data and no "ground truth" can be obtained through averaging, we evaluated the method qualitatively by visually inspecting individual spectra before and after denoising.

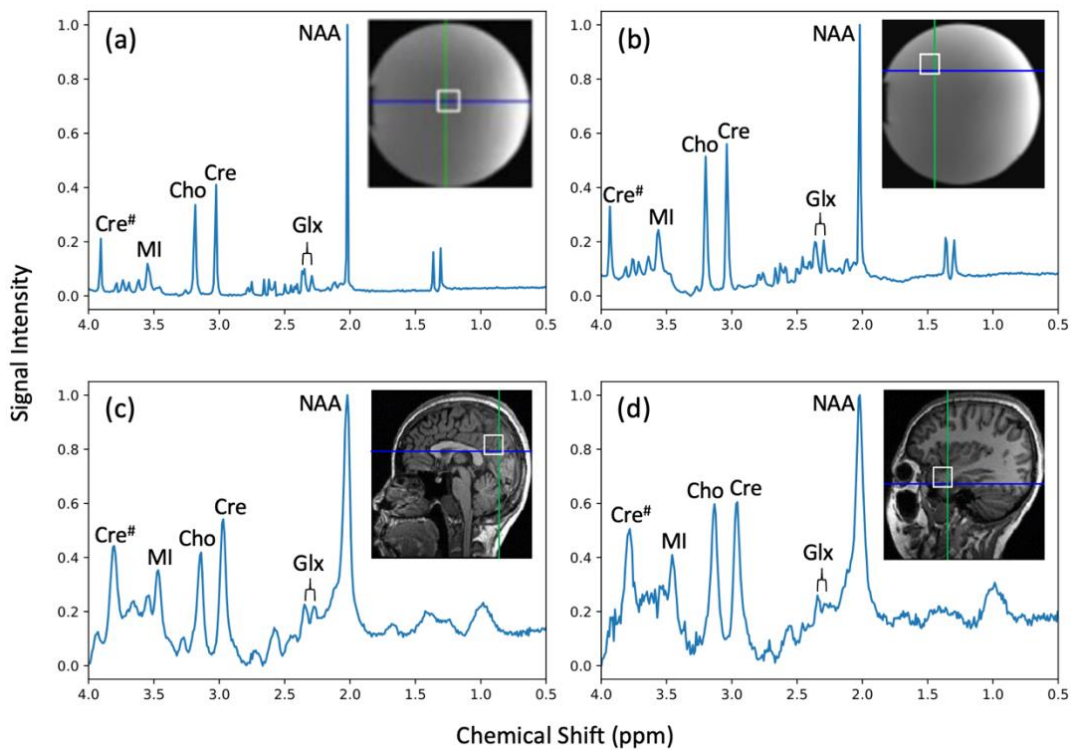

**Figure 1**. The example of the locations used for collecting MRS data (NSA = 192) from the brain phantom at center (a) and anterior high (b) locations, parietal (c) and temporal (d) regions of the human brain. Chemical shifts denoting Cre[#], MI, Cho, Cre, Glx, and NAA peaks are labeled accordingly. Glx is composed of Glu and Gln potentially overlapping at the field strength of 3 Tesla. Abbreviations: Cre[#] and Cre, creatine; # indicates the methylene group from Cre; MI, myo-inositol; Cho, choline; NAA, N-acetylaspartate. Glu, glutamate and Gln, glutamine.

**2.2. Pre-processing of MRS data**

The raw MRS data was pre-processed with following steps before being prepared as direct input patches and learning targets for SAEs. First, the raw time domain free induction decay (FID) data were transformed to the frequency domain spectra by applying Fourier transformation using LCModel (version 6.3-1H)[34]. This is followed by phase and baseline correction by LCModel[35-37]. The phase and baseline corrected spectra were then normalized as described in the previous study[33].

After preprocessing the high NSA data (NSA = 192) collected at each location, 160 out of 192 individual spectra from each high NSA data were randomly chosen and averaged to generate the high SNR "ground truth" data. For generating noise-containing input data, we randomly selected of 1, 2, 4, 8, or 16 spectra from the 160 individual spectra to make low NSA and low SNR data equivalent to those NSA of 1, 2, 4, 8, or 16 without actually acquiring them. This process was repeated 100 times for each location to obtain 100 pairs of high SNR "pure signal" ground truth and noise-containing inputs for training or evaluation purposes. The rationale and process of generating training and testing MRS dataset, e.g., NSA = 8, from high SNR data is presented in **Figure 2**. However, for the low SNR data (CSI data) collected from brain tumor patients, only the individual spectra with NSA = 1 were obtained, and there was no corresponding "ground truth" data available.

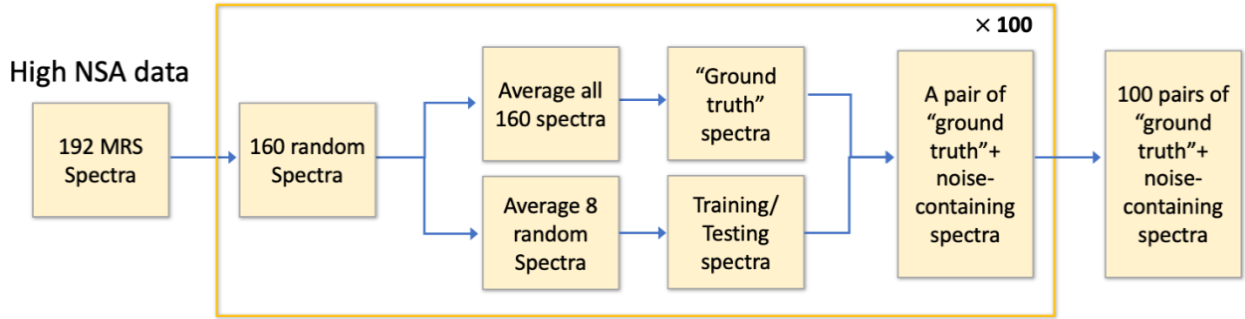

**Figure 2**. The schematic structure for creating 100 pairs of "ground truth" and noise-containing spectra from the high SNR data for training or testing purposes. The noise-free "ground truth" spectra were obtained by averaging 160 out of 192 individual (NSA of 1) spectra, while the noise-containing spectra only averaged over 1, 2, 4, or 8 individual spectra. This process was repeated 100 times for each location to obtain a diverse set of training and testing data.

**2.3. Denoising workflow**

Among the eight selected locations on the brain phantom, we used the high SNR data collected at six locations (i.e., A_DN, AH, FL, H, PL, and PS) for model training, while the high SNR data collected at the FR and PR locations from the phantom, as well as those from the temporal and parietal lobes of healthy subjects, were used as the sampling data for quantitatively evaluating the reported method. Voxel-wised spectral data extracted from clinical CSI (NSA = 1) of brain tumor patients collected previously were used to test the spectral improvement after denoising.

The main denoising workflow, as illustrated in **Figure 3**, consists of training and testing phases. In the training phase (**Figure 3(a)**), low NSA spectra and their corresponding noise-free "ground truth" were paired as the noise-containing spectra and pure signals to train the SAE models. In the testing phase (**Figure 3(b)**), the trained SAE models were applied to the input noise-containing data. The denoising performance of the method was assessed by comparing the similarity between denoised and noise-free "ground truth" signals. Metrics for quantifying the denoising performance include SNR, mean squared error (MSE), and the chemical shift value of the selected metabolite, i.e., NAA. The feed forward SAE network is depicted in **Figure 3(c)**, and the network parameters were optimized through the backward path via minimizing the loss function defined between the input and learning target.

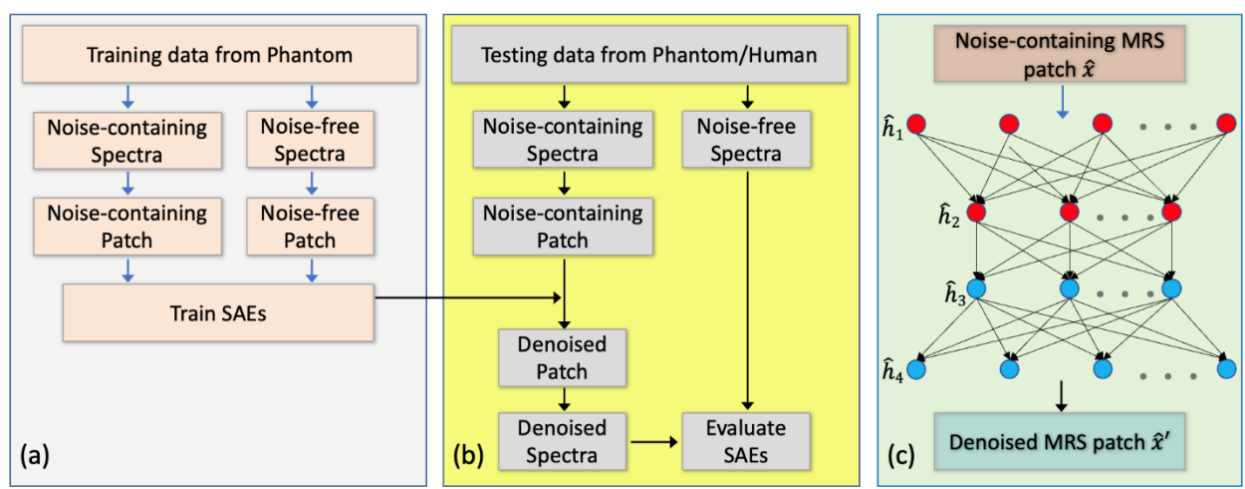

**Figure 3**. The workflow for denoising MRS spectra. (a) Training and (b) testing phases, and the sketched architecture of SAEs (c).

During the training phase, we utilized both self-supervised and fully-supervised learning to update the learning parameters for SAEs. Self-supervised learning, as commonly used in the literature[27,38], involves reconstructing the input noise-containing data with a denoised reconstruction net without referencing the "ground truth" noise-free signals, but using the input itself as the learning target. On the other hand, the fully-supervised training approach employs the noise-free "ground truth" as the learning target. The two approaches thus define different loss functions for training.

To overcome the potential overfitting problem and simplify the involved SAEs, we used the patch-based training method to augment our dataset as well as to shorten the input segments. Since we only had limited learning patterns from scans of the brain phantom and human subjects in the spectral range from 0.2-4.0 ppm in chemical shift, which covers most brain metabolites, this approach allowed us to increase the amount of available training data. We chose a patch size covering 0.29 ppm with an origin shifting of 0.02 ppm. As a result of the augmentation process, the size of the training dataset has increased significantly with 109,200 pairs of short patches comparing to the original dataset with only 600 pairs of full-scale noise-containing and noise-free spectra. For each pair of noise-free "ground truth" and noise-containing patches, both were further scaled to [0,1] linearly before being fed into the SAE networks for training or testing.

### 2.4. Training the stack auto-encoder (SAE)

The architectures of the SAEs as shown in **Figure 3 (c)** consists of a compressing encoder branch to extract the sparse deep features from the input noise-containing MRS patch $\hat{x}$ via encoding $\hat{x}$ to principle MRS features $\hat{h}$ in the hidden layers. In the feature domain, the informative part of MRS can be expressed by a few components compared to the entire feature vectors. The noise, typically represented as bias, may be related to non-sparse feature vectors. Sparsity can help to remove the noise. An expanding decoding branch is used to reconstruct the denoised MRS patch $\hat{x}'$, while the paired "ground truth" noise-free patch $\hat{y}$ was already prepared. By stacking the hidden layers, the original input $\hat{x}$ is encoded to higher-level abstractions. Through the sparsity of hidden layers, only the principle components in feature domain are kept and used for reconstruction, while the noisy components are removed without prior knowledge [39]. The objective of SAEs is to minimize the defined reconstruction error either between $\hat{x}'$ and $\hat{y}$ for fully-supervised learning, or between $\hat{x}'$ and $\hat{x}$ for self-supervised training strategy.

Let $\theta_i = \{w_i, b_i\}$ be the weighting and bias learnable parameters for the $i_{th}$ layer and $A(x) = 1/(1 + exp(-x))$ is a sigmoid activation function between stacking layers, the hidden layer can then be represented as

$$\hat{h}_1 = A(w_1\hat{x} + b_1),$$

(1)

$$\hat{h}_i = A(w_i'\hat{h}_{i-1} + b_i), \quad i = 2, 3, 4.$$

(2)

we define the decoder's output as follows:

$$\hat{x}' = A(w_{out}\hat{h}_4 + b_{out})$$

(3)

in which $\theta' = \{w_{out}, b_{out}\}$ is denoted as the weighting and bias parameters for the last output decoder layer.

For each denoised patch $\hat{x}'$, it is reconstructed form the noise-containing patch $\hat{x}$, and is expected to resemble either the noise-free signal patch $\hat{y}$ (fully-supervised learning) or the input $\hat{x}$ itself (self-supervised learning). Thus, for fully-supervised learning the learnable parameters are optimized by minimizing the reconstruction error:

$$\theta_i, \theta' = arg \min_{\theta_i,\theta'} \|\hat{x}' - \hat{y}\|_2^2,$$

( 4 )

or for self-supervised learning the learnable parameters are optimized by minimizing the reconstruction error:

$$\theta_i, \theta' = arg \min_{\theta_i,\theta'} \|\hat{x}' - \hat{x}\|_2^2.$$

( 5 )

where the *L2* norm is the mean square error aiming to minimize the difference between $\hat{x}'$ and the learning target. Both the fully-supervised and self-supervised learning approaches were used in the training process, and the results of both approaches will be presented. The training process used an Adam gradient descent optimizer with a learning rate of 0.002. The batch size was set to 50. The training dataset was split into training and validation sets using a 2:1 ratio with 2/3 of the data used for training and 1/3 used for validation.

Once the patch-based SAE network was fully trained, we tested the model with MRS data collected from FR and PR locations of the phantom as well as those from the parietal and temporal lobes of the human subjects. Similar to the processes used in the training phase, the testing spectra were also split into patches to predict denoised patches with the trained SAE models. The denoised patches were then merged to form the full-scale denoised spectra.

### 2.5. Evaluation of denoising performance

These denoised spectra were compared with the "ground truth" spectra to evaluate the performance of the SAE model. SNR and MSE were used as metrics for similarity evaluation between the noise-containing/denoised spectra and "ground truth":

$$SNR = 10 \log_{10} \frac{\sum_{i=1}^{n} x_i^2}{\sum_{i=1}^{n} (s_i - x_i)^2},$$

( 6 )

$$MSE = \frac{1}{n} \sum_{i=1}^{n} (s_i - x_i)^2,$$

( 7 )

where $x_i$ denotes the noise-containing or denoised MRS spectra, and $s_i$ is the noise-free "ground truth." Since SNR and MSE were computed for both noise-containing and denoised spectrum, we can evaluate the efficacy of SAEs by comparing the metrics before and after denoising. Additionally, to assess the ability of the denoising method to preserve metabolic information, the distinct chemical shift value of NAA peak at 2.02 ppm were identified and used to examine any distortion in the noise-containing and denoised spectra comparing to that of the "ground truth" spectra. The position change of the chemical shift of NAA in reference to the "ground truth" spectra before and after denoising can be used to evaluate whether the reported denoising method cause any displacement in the chemical shift of NAA signal.

$$\text{Shift-error}_{noise\text{-}containing} = |argmax(\hat{x}) - argmax(\hat{y})|, \tag{8}$$

$$\text{Shift-error}_{denoised} = |argmax(\hat{x}') - argmax(\hat{y})| \tag{9}$$

where $argmax(\hat{x})$, $argmax(\hat{x}')$ and $argmax(\hat{y})$ are the NAA peak positions of noise-containing inputs, denoised spectra and noise-free ("ground truth") spectra, respectively. By comparing Shift-error$_{noise\text{-}containing}$ and Shift-error$_{denoised}$ we can estimate the "drift" in the chemical shift of NAA peaks between the ground truth and the noise-containing/denoised spectra.

In addition to quantitative evaluation, the reported denoising methods were also applied to denoise CSI data, which used NSA = 1, from two brain tumor patients. Since no "ground truth" can be obtained by averaging high NSA spectra, numerical metrics such as SNR or MSE were not feasible for evaluating denoising efficacy. Instead, the denoising effect on these low SNR data was evaluated qualitatively through visual comparison between the noise-containing and denoised spectra.

## 3. RESULTS

Multiple experiments were conducted to denoise low SNR spectra with varying NSA values of 1, 2, 4, 8, and 16. The results showed that the reported network was effective in removing noise from low NSA MRS data while preserving the primary metabolite signals for quantification. **Table 1** summarizes the results from quantitative evaluation of the fully-supervised learning approach based on the measurements of SNR and MSE before and after denoising data collected from the phantom (FR location). The SNR increased and MSE decreased after denoising, demonstrating SNR enhancement. Moreover, the SNR improvement is more significant with data from lower NSA. For example, SNR increase is the greatest for NSA = 1, with 40% better while the MSE decreased 69.8%. In comparison, only 1.4% increase in SNR with only 8.3% MSE decrease was observed in denoising less noisy data from NSA = 16. This difference in performance may be because the NSA = 16 inputs are averaged over collection of "single shot" (NSA=1) of spectra with cancellation of random noise from individual spectra.

**Table 1**. Comparison of SNR and MSE values of phantom MRS spectra before and after denoising using the fully-supervised learning strategy.

| NSA | SNR (dB) | | | MSE | | |
|---|---|---|---|---|---|---|
| | Before | After | Change (%) | Before | After | Change (%) |
| 1 | 12.4 | 17.3 | 40.0 | 5.1E-03 | 1.6E-03 | -69.8 |
| 2 | 15.2 | 19.4 | 27.4 | 2.6E-03 | 9.5E-04 | -63.2 |
| 4 | 18.2 | 21.6 | 18.5 | 1.3E-03 | 5.7E-04 | -54.9 |
| 8 | 21.4 | 23.6 | 10.4 | 6.1E-04 | 3.6E-04 | -41.2 |
| 16 | 24.6 | 24.9 | 1.4 | 2.9E-04 | 2.7E-04 | -8.3 |

Note: The data were from FR location of the phantom.

From the plotted noise-containing, "ground truth", and denoised spectra, we observed improved SNR and quality of denoised spectra. **Figure 4** depicts the quality of denoising noise-containing spectra with NSA of 2 and 8 collected from the phantom (FR location) using both fully-supervised and self-supervised training approaches. In both cases, the denoised spectra are significantly smoother compared to the original noise-containing inputs, and are very close to the "ground truth". As noisy components were removed, low concentration metabolite signals emerged at distinctive chemical shifts and became detectable. Additionally, the SAE denoising models exhibited a greater smoothing effect on the noise-containing data collected with NSA = 2 compared to those with NSA = 8, despite the input data having a worse SNR.

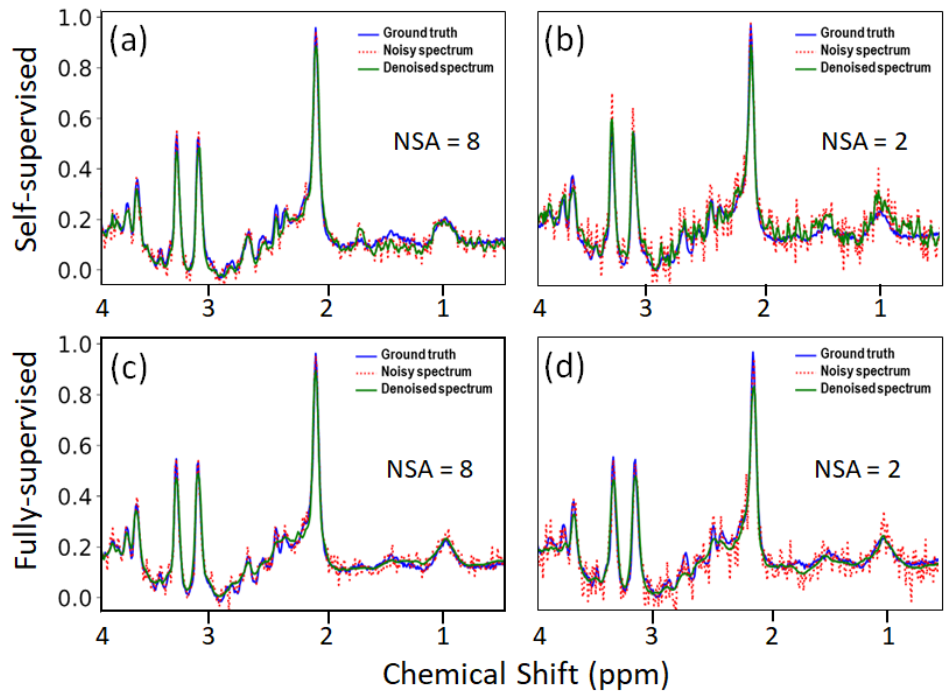

**Figure 4**. Examples of MRS spectra before and after being denoised with different approaches. Spectra were collected from the FR location of brain phantom with NSA of 2 or 8. (a, b): Denoised with self-supervised learning, and (c, d): with fully-supervised learning.

When we compared the effectiveness of denoising using different training approaches, i.e., self-supervised or fully-supervised SAEs, both training approaches achieved substantial SNR improvement and MSE reduction, yielding high quality denoised spectra. As summarized in **Table 2**, the fully-supervised learning approach outperformed the self-supervised method. In all experiments of denoising noise-containing spectra with NSA = 1, 2, 4, and 8, whether they were from the phantom or human brain, the fully-supervised models yielded a greater improvement in SNR and decrease in MSE, compared to their self-supervised counterparts. For example, training with fully-supervised SAEs resulted 43.8% SNR increase and in 68.8% MSE reduction in the denoised low SNR spectra collected from the healthy subjects with NSA of 1. When using self-supervised approach to denoise the same data, we only observed 25.7% SNR improvement with 50.0 % MSE reduction. Similarly, reported denoising approaches, both self-supervised and fully-supervised, are more effective in denoising noisier and lower NSA data than less noisy and high NSA spectra.

**Table 2**. The denoising performance of self- and fully-supervised learning as measured by SNR and MSE changes after denoising the data with different noise level and NSA; as well as the comparison of the denoising caused spectral distortion based on chemical shift "drift" of the NAA peak at 2.02 ppm.

| | Phantom | | | | | | | |
|---|---|---|---|---|---|---|---|---|
| | SNR change (%) | | MSE change (%) | | Shift "drift" ($10^{-3}$ ppm) | | | |
| NSA | Self-super-vised | Fully-super-vised | Self-super-vised | Fully-super-vised | Self-supervised | | Fully-supervised | |
| | | | | | Before | After | Before | After |
| 1 | 11.5 | 22.8 | -28.5 | -47.3 | 3.7 ± 5.1 | 3.9 ± 5.4 | 4.0 ± 5.2 | 3.7 ± 5.0 |
| 2 | 10.1 | 14.7 | -29.5 | -39.9 | 3.2 ± 4.9 | 2.6 ± 4.5 | 3.1 ± 4.7 | 2.4 ± 4.4 |
| 4 | 6.6 | 10.7 | -23.8 | -35.2 | 2.0 ± 4.1 | 2.4 ± 4.4 | 2.2 ± 4.2 | 0.8 ± 2.8 |
| 8 | 4.6 | 6.4 | -20.4 | -26.9 | 1.1 ± 3.2 | 1.7 ± 3.9 | 1.5 ± 3.6 | 0.9 ± 3.0 |
| 16 | 2.5 | 2.3 | -12.9 | -11.8 | 0.4 ± 1.9 | 1.2 ± 3.3 | 0.3 ± 1.6 | 0.9 ± 2.9 |
| **Healthy subjects** | | | | | | | | |
| 1 | 25.7 | 43.8 | -50.0 | -68.8 | 156.6 ± 456.9 | 34.5 ± 177.5 | 173.5 ± 485.5 | 26.0 ± 122.9 |
| 2 | 21.1 | 31.1 | -50.0 | -64.1 | 13.8 ± 100.8 | 7.9 ± 34.1 | 12.3 ± 79.5 | 7.6 ± 47.6 |
| 4 | 16.5 | 23.6 | -47.7 | -60.4 | 5.2 ± 5.4 | 4.4 ± 5.3 | 5.6 ± 5.5 | 3.7 ± 5.1 |
| 8 | 13.1 | 15.8 | -45.4 | -51.9 | 4.3 ± 4.9 | 3.5 ± 4.6 | 4.5 ± 5.1 | 2.7 ± 4.3 |
| 16 | 9.5 | 8.9 | -39.9 | -37.9 | 3.8 ± 4.8 | 2.8 ± 4.4 | 3.5 ± 4.8 | 2.5 ± 4.1 |

Worth noting, fully-supervised learning also resulted in less error ("drift") or distortion to the chemical shift of the metabolites, as indicated by the position of the NAA peak in comparison with that of "ground truth" spectra. As shown in **Table 2**, the spectra with lower NSA appear to have higher chemical shift "drift" comparing to the high NSA spectra before and after denoising. For example, before denoising data from healthy human, differences in the chemical shift (shift "drift") of NAA between "ground truth" spectra and NSA = 1 and NSA = 8 are 0.17 ± 0.49 ppm and 0.0045 ± 0.0051 ppm, respectively. After fully-supervised denoising, the shift "drift" decrease to 0.03 ± 0.12 ppm for the spectra with NSA = 1 and 0.0027 ± 0.0043 ppm with NSA = 8. This is because the low NSA spectra contain less information and may result in less accurate quantification of chemical shifts for the designated metabolites. Both fully-supervised and self-supervised denoising approaches yielded almost negligible "drift" in chemical shift of NAA (~ $10^{-3}$ ppm) when applied to all phantom data (NSA = 1 to 16 randomly selected from high SNR data) and most of the healthy subject data (NSA = 2 to 16 randomly selected from high SNR data). In the case of the data with NSA = 1 collected from human subjects, the chemical shift error decreased from about 0.16 to 0.03 ppm with self-supervised denoising, and it decreased from about 0.17 to 0.03 ppm with fully-supervised denoising. Both approaches showing a noticeable improvement of shift "drift" after denoising for the spectra with lower NSA. These results demonstrate that denoising with the reported deep learning methods do not cause distortion to the chemical shift of the metabolites.

In addition to the quantitative evaluations, we tested the SAEs to denoise the real-world clinical data from two brain tumor patients, retrospectively. The voxel-wise MRS data extracted from CSI were considered as low NSA noisy input data, while SAEs were trained with the phantom data. As the examples presented in **Figure 5**, the reported SAEs improved SNR of the spectra extracted from both tumor and normal regions.

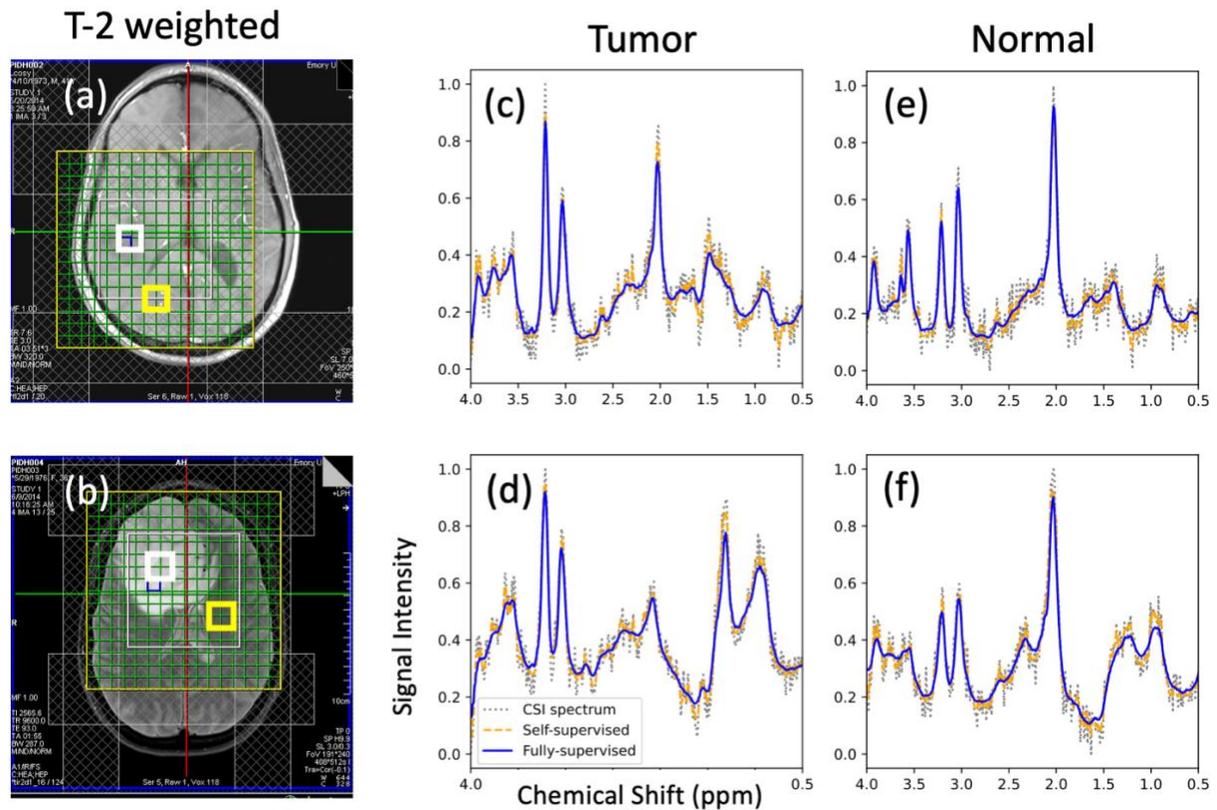

**Figure 5**. Example of denoising the low SNR spectra (NSA = 1) extracted from chemical shift images (CSI) obtained from two brain tumor patients. The regions of interest (ROIs) outlined in white are tumor (a, b), respectively, while the yellow ROIs indicate selected voxels from the normal regions. For each ROI, we present individual spectra extracted from selected voxels of CSI. The original (grey dotted lines) and denoised spectra from the tumor region (c, d) and the normal region (e, f), respectively. Both fully-supervised (blue solid lines) and self-supervised learning (orange dashed lines) SAEs were employed for denoising.

## 4. DISCUSSION

In this study, we have shown a highly effective deep learning technique for denoising MRS data without compromising the sensitivity to metabolites. Reported SAEs denoising method performed well across diverse testing datasets, including those from the standard spectroscopy phantom, healthy human brain and brain tumors in patients, demonstrating the robustness and generalizability of the models. These findings highlight the potential of SAEs as a promising tool for enhancing SNR and quality of MRS data through denoising, which can potentially impact clinical MRS applications through improving accuracy and reliability of MRS data, accelerating MRS data acquisition and streamlining clinical workflows of MRS protocols.

We have presented both fully-supervised and self-supervised learning approaches and evaluated their performances. Interestingly, both approaches are more effective in enhancing the noisier data collected with lower NSA, especially NSA = 1, compared to less noisy and averaged data collected with higher NSA (e.g., NSA = 16). It is likely that less noisy data were obtained by averaging multiple spectra in which noise cancellation takes place to those random noise components when data are averaged. While signal averaging is a classic approach to improve the SNR, it requires a longer acquisition time. Thus, high NSA data acquisition is prohibitive to MRS applications when fast data acquisition is needed in clinical situations when limited time is allowed to collect data from patients, such as patients with stroke, or data collection in the regions is motion-sensitive, such as liver or heart or even lung. Therefore, the robust ability of our denoising method to improve the SNR of extremely low NSA data, e.g., NSA = 4 or even 1, is potentially "game-changing" in clinical settings, enabling MRS applications in clinical problems that need to obtain molecular and metabolic information.

Our study has shown that, in general, fully-supervised SAEs outperforms conventional self-supervised SAEs, particularly when denoising low NSA data. This difference in denoising effectiveness may be explained by the fact that the two methods have different training targets. SAEs are commonly used in self-supervised mode, without the need for extra labeling effort[27,40-42]. They take in noise-containing input to learn high-level abstract features and then reconstruct a denoised spectrum that resembles the input itself. In other words, self-supervised SAEs have no knowledge of what a "true" spectrum should look like, but can only estimate it from the noise-containing input. For data with a higher level of noise, the estimation by the network can be substantially distracted by the embedded noise, causing errors and large deviations from the ground truth. In comparison, our proposed fully-supervised SAEs learned the "ground truth" spectra of high SNR, thus the denoising method was guided with much more information to generate the output spectra. Therefore, fully-supervised SAEs performed better in most cases. For NSA = 8 or 16 spectra, where the input itself has enough signals or features to process from multiple averages, the output spectrum is easier to reconstruct, and thus both training strategies showed comparable performance.

As a proof-of concept study, this work is limited by the data size and sources, which is a common issue in deep learning methods. In the study, we only used one type of brain phantom that contains a limited number of metabolites, leading to less diverse and complicated MRS patterns compared to those in humans. For healthy brain data, we only collected two sampling regions, i.e., the parietal and temporal lobes, while we understand that the brain tissue is heterogeneous in different areas and structures. Therefore, the representations of MRS patterns are limited in our datasets. To overcome these limitations, we employed a patch-based method that treated the full-scale spectra within a chemical shift range of 0.2-4 ppm as 182 short patches of size 0.29 ppm. By covering such a narrow spectral range, the model's complexity needed to learn high-level abstractions is largely reduced. Additionally, we augmented the dataset by generating 109,200 short patch pairs from 600 full-scale spectra pairs, thereby avoiding potential overfitting problem.

While using the spectroscopic phantom allow us to obtain "ground truth" as close as possible as we showed in our previous work[33], it is almost impossible to obtain similar "ground truth" data from human brains and patient data at this point. To overcome this challenge, we adopted a method in our study where we used high SNR spectra as the noise-free "ground truth" for both training and evaluation. Previous experiments[43,44] have used the average of 128 repeated single measurements as the "true" information, while in our work, we opted for NSA = 160 to generate the high SNR "ground truth", which involved more averages than those experiments.

While our study did not involve metabolite quantification of human subjects as a form of training information, we found that the high NSA spectra used as the "ground truth" already contained embedded metabolite quantification information. Specifically, chemical shift peaks of NAA, MI, Cre, Cre#, Cho, and Clx were present in the "ground truth" spectra. Therefore, a denoised spectra that is identical to the "ground truth" would naturally embed the correct metabolite quantification information. Consequently, the MSE calculated relative to the "ground truth" can serve as a metric to measure the similarity between the low SNR data (before or after denoising) and the "ground truth", which implicitly reflects the accuracy of the extracted metabolite quantification information.

## 5. CONCLUSIONS

This work demonstrated that the SAE networks enable the model-free denoising of MRS spectra without prior knowledge. In addition to commonly used self-supervised training strategy, the fully-supervised learning using the averaged high NSA spectra as "ground truth" achieved better results. Additionally, more significant improvements in SNR and spectral quality were obtained with input spectra that had higher noise levels. Furthermore, reported deep learning denoising methods can potentially obtain digitally enhanced high quality spectra from the noisy data collected using much lower NSA, thereby shortening the data acquisition time. Future investigation on the diagnostic value of deep-learning enhanced spectra and subsequent development and optimization could lead to clinical translation of the deep-learning approaches in improving and broadening clinical applications of MRS.

**CONFLICT OF INTEREST**

The authors declare no conflict of interest.

**ACKNOWLEDGEMENTS**

This research is supported by the grants (R01CA215718, R56EB033332, R01CA203388 and R01EB032680) from National Institutes of Health.

**References**

1. Dienel GA. Brain glucose metabolism: Integration of energetics with function. Physiological reviews. 2018;99(1):949-1045.
2. van Ewijk PA, Schrauwen-Hinderling VB, Bekkers SC, Glatz JF, Wildberger JE, Kooi ME. MRS: a noninvasive window into cardiac metabolism. NMR in Biomedicine. 2015;28(7):747-766.
3. Valkovič L, Chmelík M, Krššák M. In-vivo 31P-MRS of skeletal muscle and liver: A way for non-invasive assessment of their metabolism. Analytical biochemistry. 2017;529:193-215.
4. Wang T, Zhu XH, Li H, et al. Noninvasive assessment of myocardial energy metabolism and dynamics using in vivo deuterium MRS imaging. Magnetic resonance in medicine. 2021;86(6):2899-2909.
5. Lingvay I, Esser V, Legendre JL, et al. Noninvasive quantification of pancreatic fat in humans. The Journal of Clinical Endocrinology & Metabolism. 2009;94(10):4070-4076.
6. Hu HH, Kim HW, Nayak KS, Goran MI. Comparison of fat–water MRI and single-voxel MRS in the assessment of hepatic and pancreatic fat fractions in humans. Obesity. 2010;18(4):841-847.
7. Utriainen M, Komu M, Vuorinen V, et al. Evaluation of brain tumor metabolism with [11C] choline PET and 1H-MRS. Journal of neuro-oncology. 2003;62(3):329-338.
8. Horská A, Barker PB. Imaging of brain tumors: MR spectroscopy and metabolic imaging. Neuroimaging Clinics. 2010;20(3):293-310.
9. Sharma U, Jagannathan NR. Metabolism of prostate cancer by magnetic resonance spectroscopy (MRS). Biophysical Reviews. 2020;12(5):1163-1173.
10. Trigui R, Mitéran J, Walker PM, Sellami L, Hamida AB. Automatic classification and localization of prostate cancer using multi-parametric MRI/MRS. Biomedical Signal Processing and Control. 2017;31:189-198.
11. Haddadin IS, McIntosh A, Meisamy S, et al. Metabolite quantification and high-field MRS in breast cancer. NMR in Biomedicine: An International Journal Devoted to the Development and Application of Magnetic Resonance In vivo. 2009;22(1):65-76.
12. Stanwell P, Gluch L, Clark D, et al. Specificity of choline metabolites for in vivo diagnosis of breast cancer using 1H MRS at 1.5 T. European radiology. 2005;15(5):1037-1043.
13. Mueller S, Trabesinger A, Boesiger P, Wieser H. Brain glutathione levels in patients with epilepsy measured by in vivo 1H-MRS. Neurology. 2001;57(8):1422-1427.
14. Simister RJ, McLean MA, Barker GJ, Duncan JS. Proton MRS reveals frontal lobe metabolite abnormalities in idiopathic generalized epilepsy. Neurology. 2003;61(7):897-902.

15. Hall EL, Stephenson MC, Price D, Morris PG. Methodology for improved detection of low concentration metabolites in MRS: optimised combination of signals from multi-element coil arrays. Neuroimage. 2014;86:35-42.
16. Löffler R, Sauter R, Kolem H, Haase A, von Kienlin M. Localized spectroscopy from anatomically matched compartments: improved sensitivity and localization for cardiac31P MRS in humans. Journal of Magnetic Resonance. 1998;134(2):287-299.
17. Snaar J, Teeuwisse W, Versluis M, et al. Improvements in high‐field localized MRS of the medial temporal lobe in humans using new deformable high‐dielectric materials. NMR in Biomedicine. 2011;24(7):873-879.
18. Considine EC. The Search for Clinically Useful Biomarkers of Complex Disease: A Data Analysis Perspective. Metabolites. 2019;9(7):126.
19. Lam F, Li Y, Guo R, Clifford B, Liang ZP. Ultrafast magnetic resonance spectroscopic imaging using SPICE with learned subspaces. Magnetic resonance in medicine. 2020;83(2):377-390.
20. Lam F, Ma C, Clifford B, Johnson CL, Liang ZP. High‐resolution 1H‐MRSI of the brain using SPICE: data acquisition and image reconstruction. Magnetic resonance in medicine. 2016;76(4):1059-1070.
21. Chang SG, Yu B, Vetterli M. Adaptive wavelet thresholding for image denoising and compression. IEEE transactions on image processing. 2000;9(9):1532-1546.
22. Abramovich F, Sapatinas T, Silverman BW. Wavelet thresholding via a Bayesian approach. Journal of the Royal Statistical Society: Series B (Statistical Methodology). 1998;60(4):725-749.
23. Donoho DL, Johnstone JM. Ideal spatial adaptation by wavelet shrinkage. biometrika. 1994;81(3):425-455.
24. Donoho DL, Johnstone IM. Adapting to unknown smoothness via wavelet shrinkage. Journal of the american statistical association. 1995;90(432):1200-1224.
25. Kaur P, Singh G, Kaur P. A review of denoising medical images using machine learning approaches. Current medical imaging. 2018;14(5):675-685.
26. Ji B, Hosseini Z, Wang L, Zhou L, Tu X, Mao H. Spectral Wavelet-feature Analysis and Classification Assisted Denoising for enhancing magnetic resonance spectroscopy. NMR in Biomedicine. 2021;34(6):e4497.
27. Vincent P, Larochelle H, Lajoie I, Bengio Y, Manzagol P-A, Bottou L. Stacked denoising autoencoders: Learning useful representations in a deep network with a local denoising criterion. Journal of machine learning research. 2010;11(12).

28. Vincent P, Larochelle H, Bengio Y, Manzagol P-A. Extracting and composing robust features with denoising autoencoders. Paper presented at: Proceedings of the 25th international conference on Machine learning2008.
29. Nurmaini S, Darmawahyuni A, Sakti Mukti AN, Rachmatullah MN, Firdaus F, Tutuko B. Deep learning-based stacked denoising and autoencoder for ECG heartbeat classification. Electronics. 2020;9(1):135.
30. Xiong P, Wang H, Liu M, Lin F, Hou Z, Liu X. A stacked contractive denoising auto-encoder for ECG signal denoising. Physiological measurement. 2016;37(12):2214.
31. Xiong P, Wang H, Liu M, Liu X. Denoising autoencoder for eletrocardiogram signal enhancement. Journal of Medical Imaging and Health Informatics. 2015;5(8):1804-1810.
32. Liu X, Wang H, Li Z, Qin L. Deep learning in ECG diagnosis: A review. Knowledge-Based Systems. 2021;227:107187.
33. Lei Y, Ji B, Liu T, Curran WJ, Mao H, Yang X. Deep learning-based denoising for magnetic resonance spectroscopy signals. Paper presented at: Medical Imaging 2021: Biomedical Applications in Molecular, Structural, and Functional Imaging2021.
34. Provencher S. LCModel & LCMgui User's Manual. LCModel Version 62–4. In. 2012.
35. Hu H, Bai J, Xia G, Zhang W, Ma Y. Improved baseline correction method based on polynomial fitting for Raman spectroscopy. Photonic Sensors. 2018;8(4):332-340.
36. Xi Y, Rocke DM. Baseline correction for NMR spectroscopic metabolomics data analysis. BMC bioinformatics. 2008;9(1):324.
37. Stanford TE, Bagley CJ, Solomon PJ. Informed baseline subtraction of proteomic mass spectrometry data aided by a novel sliding window algorithm. Proteome science. 2016;14(1):19.
38. Shamaei A, Starcukova J, Pavlova I, Starcuk Jr Z. Model‐informed unsupervised deep learning approaches to frequency and phase correction of MRS signals. Magnetic Resonance in Medicine. 2022.
39. Jirayucharoensak S, Pan-Ngum S, Israsena P. EEG-based emotion recognition using deep learning network with principal component based covariate shift adaptation. The Scientific World Journal. 2014;2014.
40. Lin F, Chen K, Wang X, Cao H, Chen D, Chen F. Denoising stacked autoencoders for transient electromagnetic signal denoising. Nonlinear Processes in Geophysics. 2019;26(1):13-23.
41. Huang W-b, Sun F-c. Building feature space of extreme learning machine with sparse denoising stacked-autoencoder. Neurocomputing. 2016;174:60-71.
42. Thirukovalluru R, Dixit S, Sevakula RK, Verma NK, Salour A. Generating feature sets for fault diagnosis using denoising stacked auto-encoder. Paper presented at: 2016 IEEE International Conference on Prognostics and Health Management (ICPHM)2016.

43. Bender* A, Auer* DP, Merl T, et al. Creatine supplementation lowers brain glutamate levels in Huntington's disease. Journal of neurology. 2005;252:36-41.
44. Near J, Edden R, Evans CJ, Paquin R, Harris A, Jezzard P. Frequency and phase drift correction of magnetic resonance spectroscopy data by spectral registration in the time domain. Magnetic resonance in medicine. 2015;73(1):44-50.